\documentclass[a4paper, twocolumn, floatfix]{revtex4-2}

\usepackage{amssymb, acronym, graphicx}
\usepackage{orcidlink, adjustbox}
\usepackage{lipsum, ulem}
\usepackage{graphics}
\usepackage{amsmath, amssymb, amsfonts, placeins}
\usepackage[final]{changes} 
\usepackage{color,soul}
\usepackage[capitalise]{cleveref}
\usepackage{siunitx}
\usepackage[caption=false]{subfig}

\DeclareSIUnit{\dB}{dB}

\newacro{OPO}{optical parametric oscillator}
\newacro{OPA}{optical parametric amplifier}
\newacro{SNR}{signal-to-noise ratio}
\newacro{GW}{gravitational wave}

\begin{document}

\title{Amplified Squeezed States: Analyzing Loss and Phase Noise}

\author{K. M. Kwan\orcidlink{0009-0005-7151-4839}}%https://orcid.org/0009-0005-7151-4839
\author{M. J. Yap\orcidlink{0000-0002-6492-9156}} %https://orcid.org/0000-0002-6492-9156
\author{J. Qin\orcidlink{0000-0002-7120-9026}} %https://orcid.org/0000-0002-7120-9026
\author{D. W. Gould\orcidlink{0000-0001-8338-4289}} %https://orcid.org/0000-0001-8338-4289
\author{S. S. Y. Chua\orcidlink{0000-0001-8026-7597}} %https://orcid.org/0000-0001-8026-7597
\author{J. Junker\orcidlink{0000-0002-3051-4374}} %https://orcid.org/0000-0002-3051-4374
\affiliation{OzGrav, Centre for Gravitational Astrophysics, Research School of Physics \& Research School of Astronomy and Astrophysics, The Australian National University, Acton, Canberra, Australian Capital Territory, 2601, Australia}
\author{V. B. Adya\orcidlink{0000-0003-4955-6280}} %https://orcid.org/0000-0003-4955-6280
\affiliation{Department of Applied Physics, KTH Royal Institute of Technology, Roslagstullsbackenn 21, Stockholm SE-106 91, Sweden}
\author{T. G. McRae\orcidlink{0000-0002-6540-6824}} %https://orcid.org/0000-0002-6540-6824
\author{B. J. J. Slagmolen\orcidlink{0000-0002-2471-3828}} %https://orcid.org/0000-0002-2471-3828
\author{D. E. McClelland\orcidlink{0000-0001-6210-5842}} %https://orcid.org/0000-0001-6210-5842
\affiliation{OzGrav, Centre for Gravitational Astrophysics, Research School of Physics \& Research School of Astronomy and Astrophysics, The Australian National University, Acton, Canberra, Australian Capital Territory, 2601, Australia}

\begin{abstract}
Phase-sensitive amplification of squeezed states is a technique to mitigate high detection loss, which is especially attractive at \SI{2}{\micro\metre} wavelengths. We derived an analytical model proving that amplified squeezed states can mitigate phase noise significantly. Our model discloses two practical parameters: the effective measurable squeezing and the effective detection efficiency of amplified squeezed states. A realistic case study includes the dynamics of the gain-dependent impedance matching conditions of the amplifier. Our results recommend operating the optical parametric amplifier at high gains because of the signal-to-noise ratio’s robustness to phase noise. Amplified squeezed states are relevant in proposed gravitational wave detectors and interesting for applications in quantum systems degraded by the output coupling loss in optical waveguides.
\end{abstract}


\maketitle

\section{Introduction}

Quantum squeezed states of light are applied in diverse areas such as \ac{GW} detection \cite{LIGO_squeezed_light_application,PhysRevLett.123.231108,PhysRevLett.123.231107, LIGO2013_squeezing,Quantum_correlations,PhysRevLett.126.041102, Minjet_radiationpressure}, satellite to ground quantum key distribution \cite{Satellite_to_groundQKD}, biosensing \cite{Bowen_microscopy}, and all-optical quantum computation \cite{fewcycle_nehra,ultra_fast_computation_takanashi2020all}. A common limitation to the applications of squeezed light is the degradation of the \ac{SNR} when the squeezed light encounters optical loss during the measurement process. These losses come in the form of, for example, photodiode quantum efficiency, spatial mode mismatch, optical scattering, and absorption.

The phase-sensitive amplification of squeezed states was originally proposed by Caves \cite{Caves_quantumnoise}. 
It has been shown to mitigate detection losses in a proof of concept demonstrated by Manceau et al. \cite{manceau_su11} and has been extended to the homodyne measurement technique across an arbitrary optical bandwidth \cite{Shaked_lifting_bandwidth_limit}. Phase-sensitive amplification has also been demonstrated in waveguides to reduce the effect of outcoupling loss \cite{fewcycle_nehra} and compensate for the relatively low photodetection quantum efficiency observed at telecommunication wavelengths at THz bandwidths \cite{ultra_fast_computation_takanashi2020all}. This approach has been explored with many-body entangled states \cite{Time-reversal-based} and added to a linear interferometer's output to reduce detection loss \cite{Frascella_loss}. Phase-sensitive amplification has recently been proposed for the sub-shot noise imaging to improve its tolerance to detection loss \cite{Knyazev_absorption}.

Putting a phase-sensitive amplifier in each arm of a \ac{GW} detector was proposed \cite{PhysRevA.85.023815} to enhance the \ac{GW} signals based on the Caves model \cite{Caves_quantumnoise}.
This ``internal squeezing" concept has been refined with several configurations proposed to improve the sensitivity-bandwidth product in \ac{GW} interferometers \cite{Vaishali_internalsqueezing,Kentaro,kentarostiffbar,Korobkoquantumexpander}. Non-classical correlations generated by internal squeezing directly inside an interferometer have been demonstrated recently \cite{PhysRevLett.118.143601}. The internal squeezing approach has been further generalized and enhanced by the addition of squeezing external to the interferometric sensor. The \ac{SNR} is optimal when the internal squeezer is used as an amplifier, and the phase noise of the external squeezer is kept to a minimum \cite{korobko2023mitigating}. 

Phase noise has been observed to degrade the squeezing level after the amplification. However, the phase noise level for amplified states has neither been quantified \cite{fewcycle_nehra} nor analytically analyzed. Here, we present a model that predicts the amount of losses and phase noise in the various segments of the measurement scheme.

In this paper, we model the phase-sensitive amplification of squeezed states to mitigate detection losses after the sensor, e.g. a \ac{GW} detector. We study an \ac{OPA} seeded with a squeezed state generated from an \ac{OPO}. Our model incorporates phase noise contributions alongside the different coupling channels for optical loss for the two-cavity system. We found the impedance matching condition of the \ac{OPA} changes under different gains. The dynamics of the impedance matching conditions are crucial to understanding the two cavity systems. Our newly derived parameters of effective measurable squeezing and effective detection efficiency of the overall amplification process allow simpler characterization of future experimental realizations. The model identifies the level of losses and phase noise in the system that prevents the measurement of high-level squeezing. We conduct simulations with realistic parameters \cite{SheonOPO, Wade_squeezing,PhysRevLett.93.161105} that are compatible with the expected high levels of injected squeezing in current and future \ac{GW} detectors. 

The concept of phase-sensitive amplification has been implemented in other contexts \cite{fewcycle_nehra,ultra_fast_computation_takanashi2020all,Shaked_lifting_bandwidth_limit,Time-reversal-based,Frascella_loss,Knyazev_absorption}. However, our parameter analysis extends to include phase noise estimates and is particularly relevant for existing long-wavelength squeezed light sources \cite{Minjet_phasecontrol}. Our findings reveal that the \ac{SNR} of the amplified state is much less sensitive to phase noise arising in and after the \ac{OPA}. 

Therefore, phase-sensitive amplification allows extra flexibility for gain optimization in the \ac{OPA}. 
Considerations introduced in this paper are timely as designs are being considered for the next generation of \ac{GW} detectors that may operate with squeezed light at wavelengths in the \SI{2}{\micro\metre} region \cite{Rana_voyager,NEMO,Einstein_telescope}. The change in wavelengths is primarily motivated by thermal noise \cite{Johannes}. However, at \SI{2}{\micro\metre} the photodiodes usually have quantum efficiencies of only \SI{74}{\percent} \cite{Georgia_squeezed_light_paper}. The best quantum efficiency measurements of photodetectors at these longer wavelengths reach only around \SI{92}{\percent} \cite{Steinlechner_squeezedlight}. 
This compares unfavorably with \ac{GW} detectors operating today at a wavelength of \SI{1064}{\nano\metre} that use photodiodes with quantum efficiencies of \SI{98}{\percent} \cite{Maggie_ligosqueezing}. In addition, experimental demonstration for a high level of squeezing at \SI{1064}{\nano\metre} was achieved for photodiodes with a quantum efficiency of \SI{99.5}{\percent} \cite{Vahlbruch_15dB}.

\section{Conventional detection of squeezed states}
\label{section: lossy_detection_of_squeezed_states}

As an introduction to our model, we review the effect of detection loss on the output variances of a squeezed state generated by an \ac{OPO}. A comprehensive review can be found in \cite{minjet_thesis2020}.

\begin{figure}[t!]
\centering
\includegraphics[width=\columnwidth]{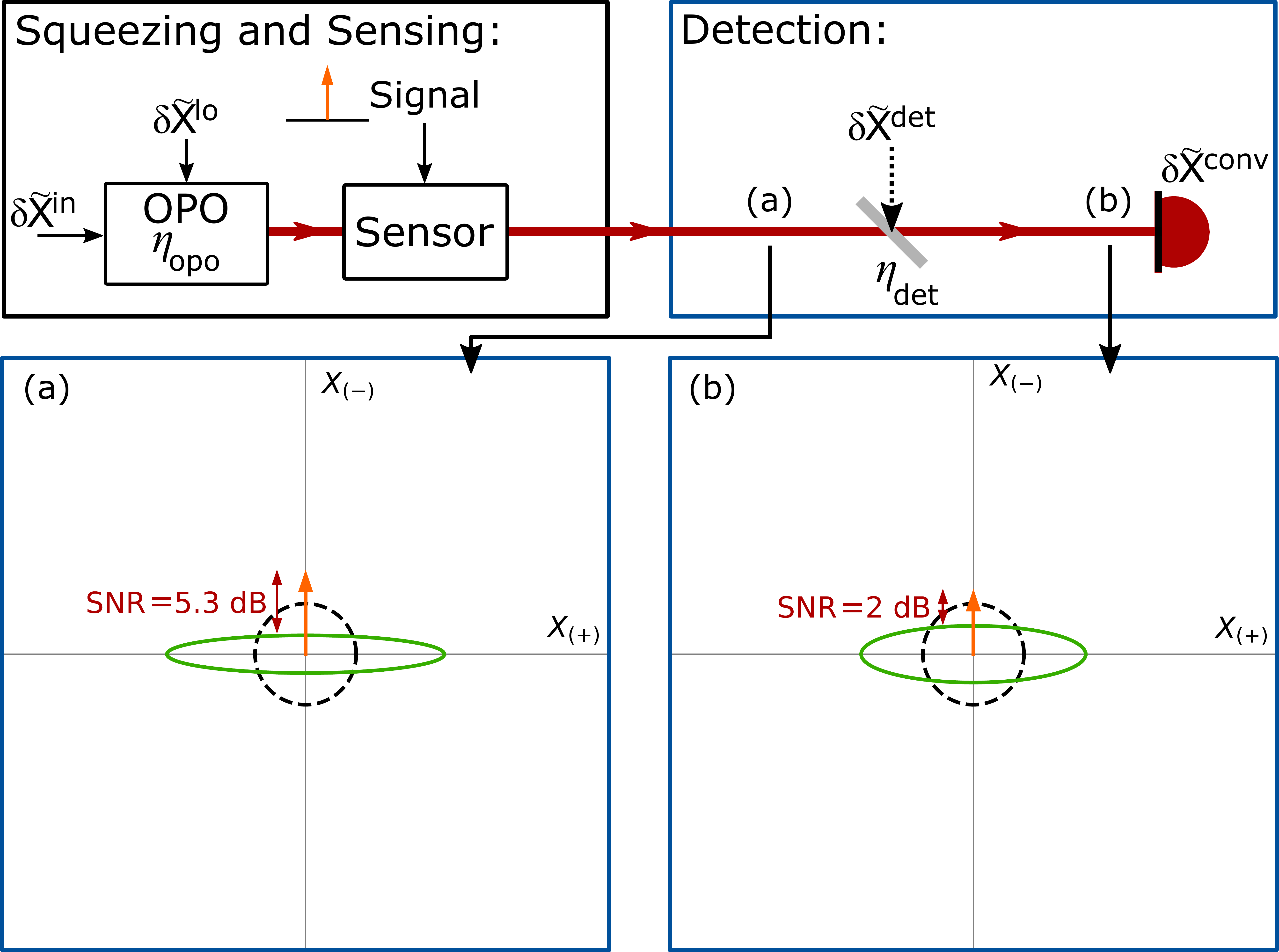}
\caption{\label{one opo block diagram}Top row: Conventional setup of employing squeezed states. The squeezed state from the \ac{OPO} collects a signal of interest at the sensor before the detection. Quadrature fluctuations couple into the setup at the \ac{OPO} ($\delta \mkern-1mu \tilde{X}^{\rm in}$, $\delta \mkern-1mu \tilde{X}^{\rm lo}$) and at the detection ($\delta \mkern-1mu \tilde{X}^{\rm det}$). Bottom row: Phase state pictures of the squeezed state at two points in the setup. (a) The state from the \ac{OPO} (gain of 1.8) is squeezed by $\SI{4.3}{\deci \bel}$ in its phase quadrature $X_{(-)}$ and senses a signal with an \ac{SNR} of \SI{5.3}{\deci \bel}. (b) The state experiences optical losses at the detection, resulting in a less squeezed state with reduced signal strength and a degraded \ac{SNR} of \SI{2}{\deci \bel}. The dashed circle shows a vacuum state uncertainty as a reference. The squeezing level was chosen for visualization purposes.}
\end{figure}

The conventional sensing and measurement scheme is shown in the upper row of \cref{one opo block diagram}, where the squeezed states are used to reduce the vacuum noise coupling into a sensor. The squeezed state is generated by a vacuum-seeded degenerate \ac{OPO} pumped by the second harmonic below oscillation threshold, which is modeled using the Hamiltonian approach for an optical resonator with a $\chi^{(2)}$ nonlinear crystal \cite{Collett_Gardiner_squeezinghamiltonian}. Losses inside the \ac{OPO} are incorporated in the vacuum field $\mathbf{A}_{\rm lo}$ coupled through the loss port into the cavity. 
The equation of motion for the intracavity field is written in a compact matrix form:
\begin{equation} 
\mathbf{\dot{a}= M}_{\rm a} \mathbf{a + M}_{\rm in} \mathbf{A}_{\rm in} + \mathbf{M}_{\rm l} \mathbf{A}_{\rm lo} , 
\label{equation fundamental equation of motion in matrix form}
\end{equation}
where $\mathbf{a}$ is the vector for the cavity mode, and ${\mathbf{A}_{\rm in}}$, ${\mathbf{A}_{\rm lo}}$ are the vectors for the fields entering the cavity via the input and loss ports respectively:
\begin{equation} 
\mathbf{\dot{a}}=  \begin{pmatrix} \dot{a} \\ \dot{a}^\dag \end{pmatrix}  , \mathbf{a} = \begin{pmatrix} a \\ a^\dag \end{pmatrix}, 
\label{equation intracavity vector matrix}
\end{equation}
\begin{equation}
\mathbf{A}_{\rm in}= \begin{pmatrix} {A_{\rm in} } \\[1ex]  {A_{\rm in}^\dag} \end{pmatrix}, \mathbf{A}_{\rm lo} = \begin{pmatrix} {A_{\rm lo} } \\[1ex] {A_{\rm lo}^ \dag} \end{pmatrix} .
\label{equation field vector matrix}
\end{equation}
The corresponding matrices of the field vectors are:
\begin{equation}  
\mathbf{M}_{\rm a}= \begin{pmatrix} -\kappa^{\rm a} & |q|e^{i\phi} \\ |q|e^{-i\phi} & -\kappa^{\rm a} \end{pmatrix} , 
\label{equation intracavity field matrix}
\end{equation}
\begin{equation} 
\mathbf{M}_{\rm in} = \sqrt{2 \kappa^{\rm a}_{\rm in}}\mathbf{I} \;,\;  \mathbf{M}_{\rm l}= \sqrt{2 \kappa^{\rm a}_{\rm l}}\mathbf{I} , 
\label{equation input loss field matrix}
\end{equation}
where $\kappa^{\rm a}_{\rm in}$ and $\kappa^{\rm a}_{\rm l}$ is the decay rate of the mirror at the input and loss port, $\kappa^{\rm a}$ is the total decay rate of the cavity and $|q|e^{\pm i\phi}$ is the nonlinear gain factor expressed in amplitude and phase components. The matrix for the intracavity field $\mathbf{M}_{\rm a}$ is simplified by assuming there are no phase differences between the pump and fundamental field, ${\phi = 0}$.

For frequencies within the cavity linewidth, the output field in the Fourier domain is: 
\begin{align} 
\tilde{\mathbf{A}}_{\rm out} = & \mathbf{M}_{\rm in} \mathbf{\tilde{a}} - \mathbf{\tilde{A}}_{\rm in} \notag \\
 =  &  \mathbf{M}_{\rm in} \left[ (i\Omega \mathbf{I- M}_{\rm a})^{-1} (\mathbf{M}_{\rm in} \mathbf{\tilde{A}}_{\rm in} +  \mathbf{M}_{\rm l} \mathbf{\tilde{A}}_{\rm lo}) \right]  \notag \\
&   - \mathbf{\tilde{A}}_{\rm in}  . 
\label{equation output field from opo}
\end{align}

The fields are linearized into a steady state component and a fluctuating component, ${\mathbf{A} =  \mathbf{\bar{A}} +\mathbf{\delta\mkern-1mu{A}} }$, where ${ \mathbf{\bar{A}}  = 0}$ for the vacuum field. \Cref{equation output field from opo} is expressed in quadrature fluctuation via the following transformation:
\begin{equation}
\delta\mkern-1mu\mathbf{\tilde{X}}^{k} = \mathbf{\Gamma} \delta \mkern-2mu \mathbf{\tilde{A}}_{k}  =  \begin{pmatrix} -i & i \\ 1 & 1 \end{pmatrix} \begin{pmatrix} \delta \mkern-1mu \tilde{A}_{k} \\[1ex] \delta \mkern-1mu \tilde{A}_{k}^\dag \end{pmatrix} = \begin{pmatrix} \delta \mkern-1mu \tilde{X}_{(-)}^{k} \\[1ex] \delta \mkern-1mu \tilde{X}_{(+)}^{k} \end{pmatrix} ,
\label{equation quadrature operator}
\end{equation}
where $\mathbf{\Gamma}$ is the conversion matrix between the field fluctuations $\delta \mkern-2mu \mathbf{\tilde{A}}_{k}$ and the quadrature fluctuations $\delta \mkern-1mu \mathbf{\tilde{X}}^{k}$, and ${k \in \{ \text{in , out , lo}\}}$ for the input, output and loss of the field entering and exiting \ac{OPO}.

The noise quadratures at the output of the \ac{OPO} are:
\begin{align} 
\delta \mkern-1mu \mathbf{\tilde{X}}^{\rm out} =  & \left( \mathbf{M}_{\rm in} \mathbf{\Gamma} (i\Omega \mathbf{I- M}_{\rm a})^{-1} \mathbf{M}_{\rm in} \mathbf{\Gamma}^{-1} - \mathbf{I} \right) \delta \mkern-1mu \mathbf{\tilde{X}}^{\rm in}   \notag \\ 
&  + \left( \mathbf{M}_{\rm in} \mathbf{\Gamma} (i\Omega \mathbf{I- M}_{\rm a})^{-1} \mathbf{M}_{\rm l} \mathbf{\Gamma}^{-1} \right) \delta \mkern-1mu \mathbf{\tilde{X}}^{\rm lo} \notag \\
= & \mathbf{M}^{\rm opo}_{\rm in} \delta \mkern-1mu \mathbf{\tilde{X}}^{\rm in} + \mathbf{M}^{\rm opo}_{\rm l} \delta \mkern-1mu \mathbf{\tilde{X}}^{\rm lo}  .
\label{equation opo output noise quadrature in unsolved matrix form}
\end{align}

Here, the noise quadrature for the output of the \ac{OPO} is written in the form of the transfer function for each port of the \ac{OPO} cavity.
The matrices of the noise quadrature at the input ($\bf{M}_{\rm in}^{\rm opo}$, defined in reflection) and the loss port ($\bf{M}_{\rm l}^{\rm opo}$, defined in transmission) of the \ac{OPO} are:
\begin{equation} 
\mathbf{M}_{\rm in}^{\rm opo} = \begin{bmatrix} \frac{2\eta_{\rm opo}}{1+x_{\rm opo}}-1 & 0 \\ 0 & \frac{2\eta_{\rm opo}}{1-x_{\rm opo}}-1 \end{bmatrix} , 
\label{opo input matrix}
\end{equation}

\begin{equation} 
\mathbf{M}_{\rm l}^{\rm opo} = \begin{bmatrix} \frac{2\sqrt{\eta_{\rm opo}(1-\eta_{\rm opo})}}{1+x_{\rm opo}} & 0 \\ 0 & \frac{2\sqrt{\eta_{\rm opo}(1-\eta_{\rm opo})}}{1-x_{opo}} \end{bmatrix} . 
\label{opo loss matrix}
\end{equation}

The main diagonal elements of the transfer matrices represent the components for the squeezed phase quadrature $\tilde{X}_{(-)}^{\rm out}$, and the anti-squeezed amplitude quadrature $\tilde{X}_{(+)}^{\rm out}$, where ${\eta_{\rm opo} = \kappa^{\rm a}_{\rm in} / \kappa^{\rm a} }$ is the escape efficiency of the \ac{OPO}. 
The normalized pump parameter is defined as ${x_{\rm opo} = |q|/ \kappa^{\rm a}  } $, which can be expressed in terms of the OPO nonlinear gain $G_{\rm opo}$ as:
\begin{equation} 
x_{\rm opo} = 1 - \frac{1} {\sqrt{G_{\rm opo}}}. 
\label{normalised pump parameter to gain}
\end{equation}

In the conventional detection scheme, losses are modeled as the coupling of a quadrature fluctuation $ \delta \mkern-1mu \mathbf{\tilde{X}}^{\rm det}$ into the output quadrature of the \ac{OPO} via a beamsplitter with detection efficiency $\eta_{\rm det}$. Note that $\eta_{\rm det}$ explicitly includes \textit{all} losses occurring during the detection process, such as propagation losses, mode-mismatch in the case of balanced detection, and the quantum efficiency of the photodiodes. Therefore, the quadrature at the photodetector in a conventional detection scheme can be written as:

\begin{align}
\delta \mkern-1mu \mathbf{\tilde{X}}^{\rm conv} & = \sqrt{\eta_{\rm det}} \delta \mkern-1mu \mathbf{\tilde{X}}^{\rm out} + \sqrt{1 -  \eta_{\rm det}} \delta \mkern-1mu \mathbf{\tilde{X}}^{\rm det}   \notag \\
& =  \mathbf{TF}_{\rm in}^{\rm conv} \delta \mkern-1mu \mathbf{\tilde{X}}^{\rm in} + \mathbf{TF}_{\rm lo}^{\rm conv} \delta \mkern-1mu \mathbf{\tilde{X}}^{\rm lo} + \mathbf{TF}^{\rm conv}_{\rm det}  \delta \mkern-1mu \mathbf{\tilde{X}}^{\rm det}.  
\label{equation output noise quadrature to include detection loss}
\end{align}
The transfer function of the quadratures coupled into the setup are:
\begin{align} 
\mathbf{TF}_{\rm in}^{\rm conv} & = \sqrt{ \eta_{\rm det}} \; \mathbf{M}_{\rm in}^{\rm opo} \; , \\  
\mathbf{TF}^{\rm conv}_{\rm lo} & = \sqrt{ \eta_{\rm det}} \; \mathbf{M}_{\rm l}^{\rm opo} \;,\\
\mathbf{{TF}}^{\rm conv}_{\rm det} & = \sqrt{1-\eta_{\rm det}} \; \mathbf{I} \;.
\label{equation: transfer function conventional}
\end{align}

The noise variances for the conventional detection scheme at the photodiode are given by:
\begin{equation} \mathbf {V}^{\rm conv} = \begin{pmatrix} V^{\rm conv}_{(-)} \\[1ex] V^{\rm conv}_{(+)}  \end{pmatrix} = \langle| \mathbf{\delta \tilde{X}}^{\rm conv}|^2\rangle = \begin{pmatrix} 1 - \frac{4x_{\rm opo} \eta_{\rm sqz} }{(1 + x_{\rm opo})^2} \\[1ex] 1 + \frac{4x_{\rm opo} \eta_{\rm sqz} }{(1 - x_{\rm opo})^2} \end{pmatrix} \; .
\label{single opo variance}
\end{equation}
For simplification, the escape efficiency $\eta_{\rm opo}$, the detection efficiency $\eta_{\rm det}$ are combined into the squeezing efficiency ${\eta_{\rm sqz} = \eta_{\rm opo} \eta_{\rm det}}$ and the mean quadrature fluctuation is ${\langle \delta \mathbf{\tilde{X}}^{\rm conv} \rangle \approx 0 }$. In \cref{single opo variance}, the elements in the matrix correspond to the squeezed and anti-squeezed quadrature at the photodiode respectively.

\begin{table}[t!]
    \centering
    \begin{tabular}{c|c}
        
        \textbf{Parameter} & \textbf{Value} \\
        \hline
        $\eta_{\rm opo}$, $\eta_{\rm opa}$ & 0.98 \\

        $\eta_{\rm prop}$ & 0.99 \\

       $\eta_{\rm det}$ & 0.7  \\

    \end{tabular}    
    \caption{\label{parameters table}Realistic parameters used, unless otherwise stated.}
    
\end{table}

Before detection, the squeezed state collects a phase signal of interest at the sensor, depicted as the orange arrow in the ${X_{(-)}\text{-direction}}$ in the diagram \cref{one opo block diagram}a. We have chosen realistic parameters listed in \cref{parameters table} and an arbitrary signal of \SI{1}{\deci\bel} above the vacuum state. The \ac{OPO} generates a \SI{4.3}{\deci\bel} squeezed state, resulting in an \ac{SNR} of \SI{5.3}{\deci\bel}. 

The effect of detection losses on the squeezed state is illustrated in \cref{one opo block diagram}b. Two relevant processes happen simultaneously but on different scales. The signal's classical amplitude reduces, and the squeezed noise becomes larger because the squeezed state is mixed with the vacuum state. Hence, the \ac{SNR} decreases to $\SI{2}{\deci \bel}$ in our example.

\section{Amplifying squeezed states to enhance the detection efficiency}
\label{section amplify}

\begin{figure}[t!]
\includegraphics[width=\columnwidth]{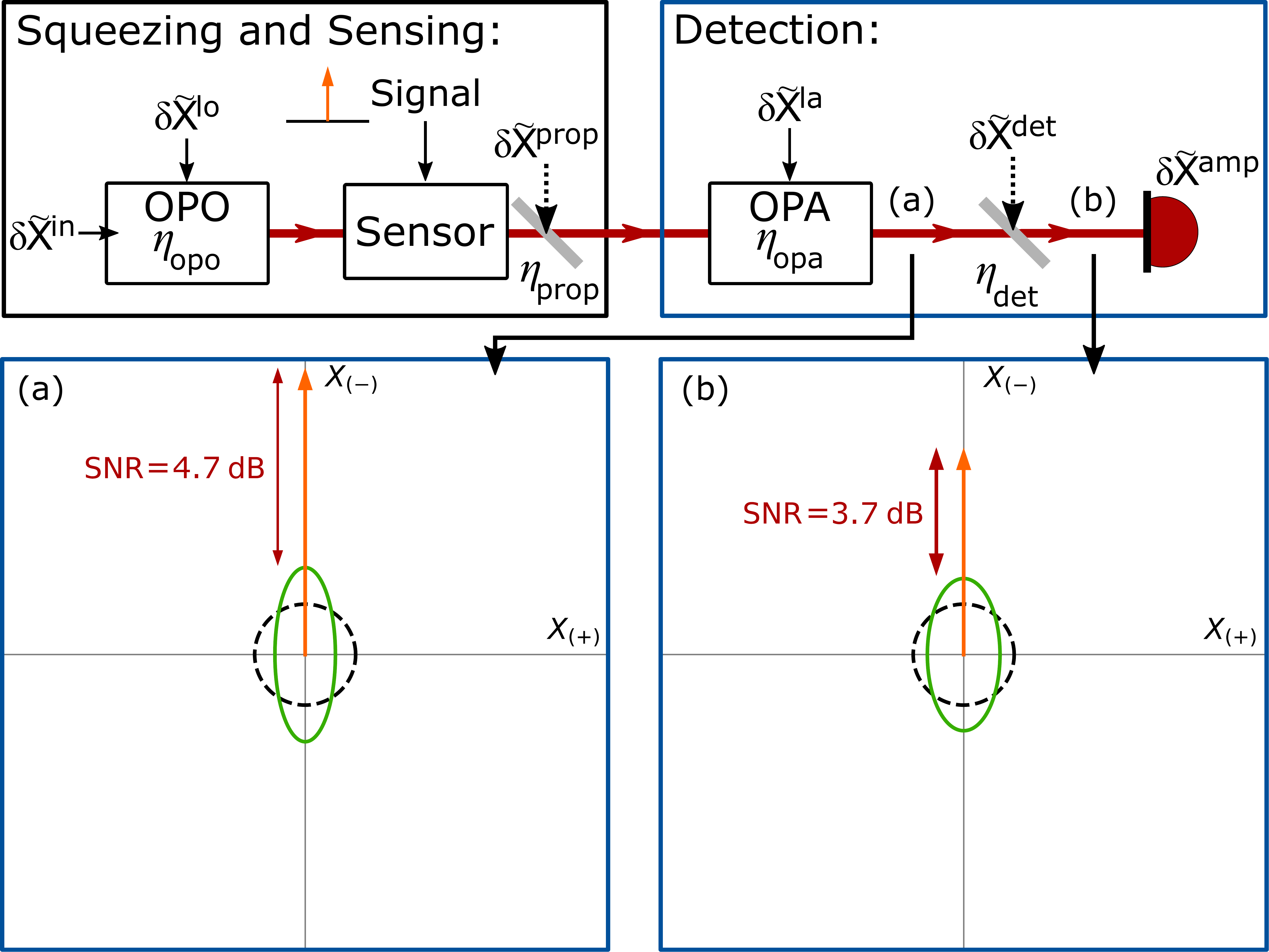} 
\caption{\label{double opo block diagram}Top row: Detection of the amplified state. The squeezed state from the \ac{OPO} is phase-sensitively amplified by the \ac{OPA} after the sensor. Bottom row: Phase state pictures of the state at two points in the setup. (a) After amplification (gain of 2.4), the state is anti-squeezed above vacuum noise, but the signal is also amplified in the phase quadrature $X_{(-)}$, leading to an \ac{SNR} of \SI{4.7}{\deci \bel}. Now, optical losses have a smaller effect on the state. (b) At the detection, the \ac{SNR} stays relatively constant and only drops by \SI{1}{\deci \bel} to \SI{3.7}{\deci \bel}. The dashed circle again shows a vacuum state as a reference.}
\end{figure}

This section shows how adding an \ac{OPA} makes the \ac{SNR} more robust to detection losses. Similar to the \ac{OPO}, we model the amplification process with a second cavity that contains a $\chi^{(2)}$ nonlinear crystal. The \ac{OPA} is placed after the sensor, see upper row of \cref{double opo block diagram}. 

The noise quadrature for the cascaded two-cavity system is calculated by: 

\begin{align}
\delta \mathbf{\tilde{X}}^{\rm amp} = & \; \mathbf{TF}_{\rm in}^{\rm amp} \delta \mathbf{\tilde{X}}^{\rm in} +\mathbf{TF}_{\rm lo}^{\rm amp}  \delta \mathbf{\tilde{X}}^{\rm lo} + \mathbf{TF}_{\rm prop}^{\rm amp}  \delta \mathbf{\tilde{X}}^{\rm prop}  \notag \\
& + \mathbf{TF}_{\rm la}^{\rm amp}  \delta \mathbf{\tilde{X}}^{\rm la} + \mathbf{TF}_{\rm det}^{\rm amp}  \delta \mathbf{\tilde{X}}^{\rm det},
\label{field equation for double opo}
\end{align}
where we include vacuum noise quadrature fluctuations from propagation loss $\delta \mathbf{\tilde{X}}^{\rm prop}$ and intracavity loss in the \ac{OPA} $\delta \mathbf{\tilde{X}}^{\rm la}$. 
The transfer matrices of the \ac{OPA} $\bf{M}_{\rm in}^{\rm opa}$ and $\bf{M}_{\rm l}^{\rm opa}$  are defined in a similar manner as for the \ac{OPO} in \cref{opo input matrix}:  
\begin{equation} 
\mathbf{M}_{\rm in}^{\rm opa} = \begin{bmatrix} \frac{2\eta_{\rm opa}}{1-x_{\rm opa}}-1 & 0 \\ 0 & \frac{2\eta_{\rm opa}}{1+x_{\rm opa}}-1 \end{bmatrix} , 
\label{opa input matrix}
\end{equation}

\begin{equation} 
\mathbf{M}_{\rm l}^{\rm opa} = \begin{bmatrix} \frac{2\sqrt{\eta_{\rm opa}(1-\eta_{\rm opa})}}{1-x_{\rm opa}} & 0 \\ 0 & \frac{2\sqrt{\eta_{\rm opa}(1-\eta_{\rm opa})}}{1+x_{\rm opa}} \end{bmatrix} . 
\label{opa loss matrix}
\end{equation}
Note that these transfer matrices are defined such that the squeezing angle of the \ac{OPA} is orthogonal to the one of the \ac{OPO}. 

The transfer functions of the quadrature fluctuation at different points of the setup are:
\begin{align} 
\mathbf{TF}^{\rm amp}_{\rm in} & = \sqrt{\eta_{\rm prop} \eta_{\rm det}} \; \mathbf{M}_{\rm in}^{\rm opa} \; \mathbf{M}_{\rm in}^{\rm opo} \; , \\  
\mathbf{TF}^{\rm amp}_{\rm lo} & = \sqrt{\eta_{\rm prop} \eta_{\rm det}} \; \mathbf{M}_{\rm in}^{\rm opa} \; \mathbf{M}_{\rm l}^{\rm opo} \;,\\
\mathbf{TF}^{\rm amp}_{\rm prop} & = \sqrt{(1-\eta_{\rm prop}) \eta_{\rm det}} \; \mathbf{M}_{\rm in}^{\rm opa} \;,\\
\mathbf{TF}^{\rm amp}_{\rm la} & = \sqrt{\eta_{\rm det}} \; \mathbf{M}_{\rm l}^{\rm opa}\; ,\\
\mathbf{TF}^{\rm amp}_{\rm det} & = \mathbf{TF}^{\rm conv}_{\rm det} = \sqrt{1-\eta_{\rm det}} \; \mathbf{I} \;.
\label{transfer function}
\end{align} 
We group all losses between the two cavities together into the propagation loss ${\eta_{\rm prop}}$. The \ac{OPA} also introduces an additional \ac{OPA} escape efficiency term $\eta_{\rm opa}$.

The noise variance of the amplified detection in the $X_{(-)}$ quadrature is calculated from \cref{field equation for double opo}: 
\begin{align}
V_{(-)}^{\rm amp} =  1 & + \frac{4x_{\rm opa}\eta_{\rm det} \eta_{\rm opa}} {(1-x_{\rm opa})^2} \notag \\
& - \frac{4x_{\rm opo} \tilde{\eta}_{\rm sqz}}{(1+x_{\rm opo})^2} \left[ \frac{\eta_{\rm det}  (2\eta_{\rm opa} + x_{\rm opa} - 1 ) ^2} { (1-x_{\rm opa})^2} \right].
\label{variance for amplified squeezing}
\end{align}

In the phase space picture, the \ac{OPA} anti-squeezes the input state, see picture \cref{double opo block diagram}a, such that both the signal and the noise are amplified in the $X_{(-)}$-direction. In \cref{variance for amplified squeezing}, the escape efficiency of the \ac{OPO} and the propagation efficiency are combined into the squeezing efficiency term, where ${\tilde{\eta}_{\rm sqz} = \eta_{\rm opo}\eta_{\rm prop}}$. Due to the non-unity efficiencies of $\tilde{\eta}_\text{sqz}$ and $\eta_\text{opa}$, the \ac{SNR} of the state after the \ac{OPA} is slightly reduced to \SI{4.7}{\deci \bel} as compared to the conventional detection case from \cref{one opo block diagram}. However, the \ac{SNR} is now much more robust to subsequent losses included in $\eta_{\rm det}$, which is depicted in \cref{double opo block diagram}b. Compared to conventional detection, the \ac{SNR} degrades less to a value of \SI{3.7}{\deci \bel}. Our model shows that increasing the amplification by the \ac{OPA} in the $X_{(-)}$-direction results in decreasing the impact of detection optical losses.

An optimal detection scheme would benefit from all of the squeezing generated by the \ac{OPO}.
We introduce the parameter of \textit{effective measurable squeezing} $V_{\rm eff}$ to quantify how much of the initial squeezing can be recovered by the detection technique. 
It is defined by referencing the squeezed state amplified by the \ac{OPA} (\cref{variance for amplified squeezing}) to a state generated with the same but vacuum seeded \ac{OPA} (\cref{variance for amplified squeezing} with ${x_{\rm opo} = 0}$):
\begin{equation}
V_{(-)}^{\rm amp} \big| _{x_{\rm opo}=0} = 1 + \frac {4x_{\rm opa}\eta_{\rm det}\eta_{\rm opa}} {(1-x_{\rm opa})^2} .
\label{variance for amplified vacuum}
\end{equation}
The effective measurable squeezing simplifies to 
\begin{equation}
V_{\rm eff}  = \frac{V_{(-)}^{\rm amp}} {V_{(-)}^{\rm amp}\big| _{x_{\rm opo}=0}} = 1 - \frac{4x_{\rm opo} \tilde{\eta}_{\rm sqz} \eta_{\rm eff}} {(1+x_{\rm opo})^2} , 
\label{effective variance for double opo}
\end{equation}
by introducing the \textit{effective detection efficiency} $\eta_{\rm eff}$. So \cref{effective variance for double opo} has the same form as the \ac{OPO} squeezing output from \cref{single opo variance}. The effective detection efficiency is:

\begin{equation} 
\eta_{\rm eff} =  \frac{\eta_{\rm det}  (2\eta_{\rm opa} + x_{\rm opa} - 1 ) ^2} { (1-x_{\rm opa})^2 + 4x_{\rm opa} \eta_{\rm det} \eta_{\rm opa}  } .
% &= \frac {\eta_{\rm det} \left( 1- \eta_{\rm opa} (\sqrt{G_{\rm opa}} +1) \right) ^2 } {1+ (G_{\rm opa}-1) \eta_{\rm det} \eta_{\rm opa}} . %removed
\label{effective efficiency}
\end{equation}
It consists of the \ac{OPA} escape efficiency $\eta_{\rm opa}$, gain $G_{\rm opa}$ and the detection efficiency $\eta_{\rm det}$. 
The gain of the \ac{OPA} determines the effective measurable squeezing level at the output of the setup. 

\begin{figure}[t!]
\centering
\includegraphics[width=0.98\columnwidth]{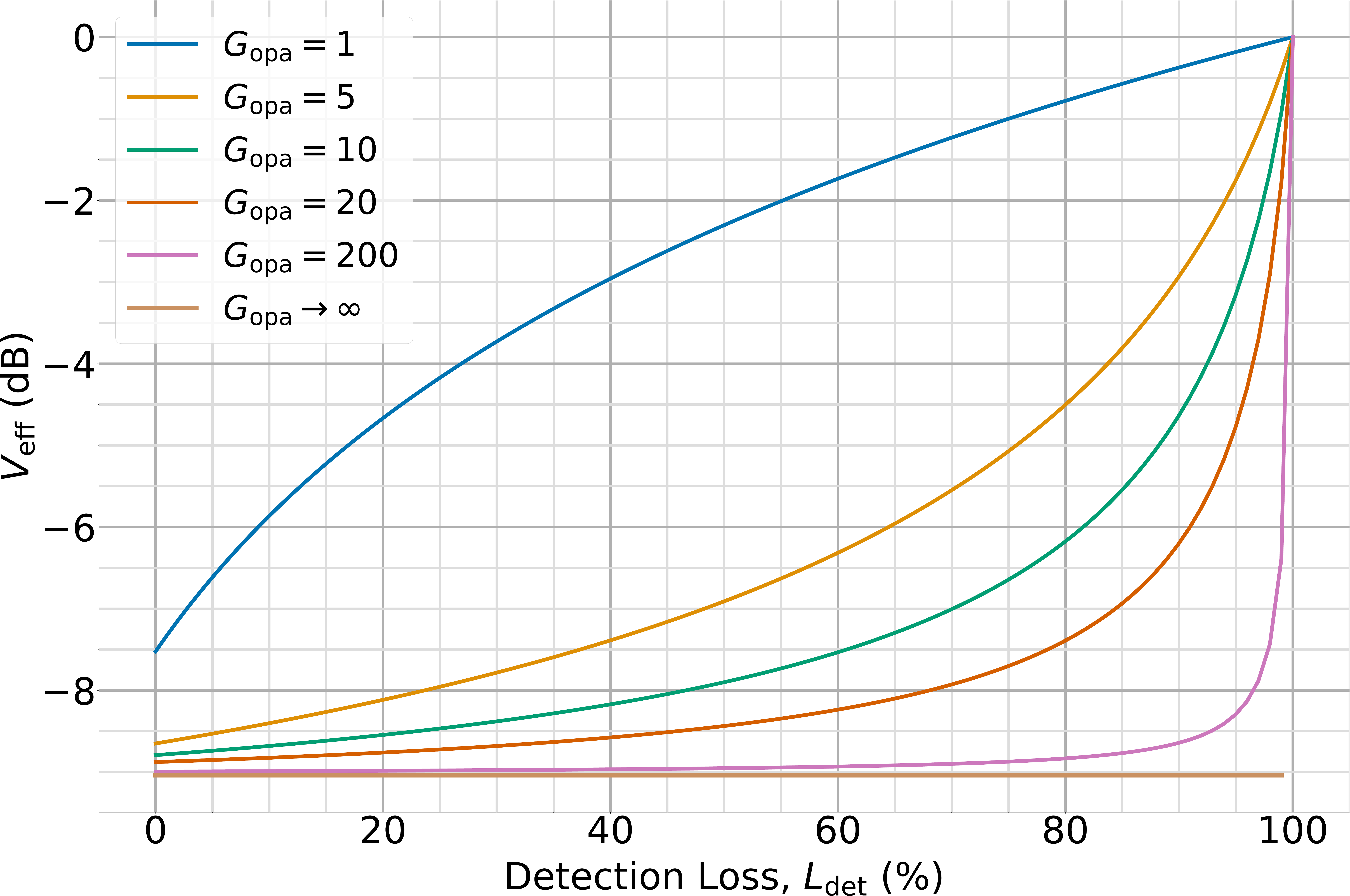}	
\caption{\label{Gain effect with lossy}The dependence of the amplified state by the \ac{OPA} on subsequent detection losses $L_{\rm det}$. The effective measurable squeezing $V_{\rm eff}$ can be recovered for high gain values $G_{\rm opa}$.}

\end{figure}

Figure \ref{Gain effect with lossy} shows how the effective measurable squeezing $V_{\rm eff}$ changes with detection losses $L_{\rm det} = 1 - \eta_{\rm det}$. The squeezed state generated by the \ac{OPO} and reaching the \ac{OPA} has a squeezing value of \SI{-9.1}{\deci \bel} (${G_{\rm opo} = 5.2}$). For even high detection losses, we can obtain large effective measurable squeezing values. The higher the gain $G_{\rm opa}$, the closer we reach the theoretical \SI{-9.1}{\deci \bel} limit. For instance, for a gain of ${G_{\rm opa }=10 }$ and detection losses of about \SI{30}{\percent}, we can still recover ${V_{\rm eff} \approx \SI{-8.4}{\deci \bel}}$.

\begin{figure}[t!]
  \subfloat[\label{neff for g2=1}]{
    \includegraphics[width=0.46\textwidth]{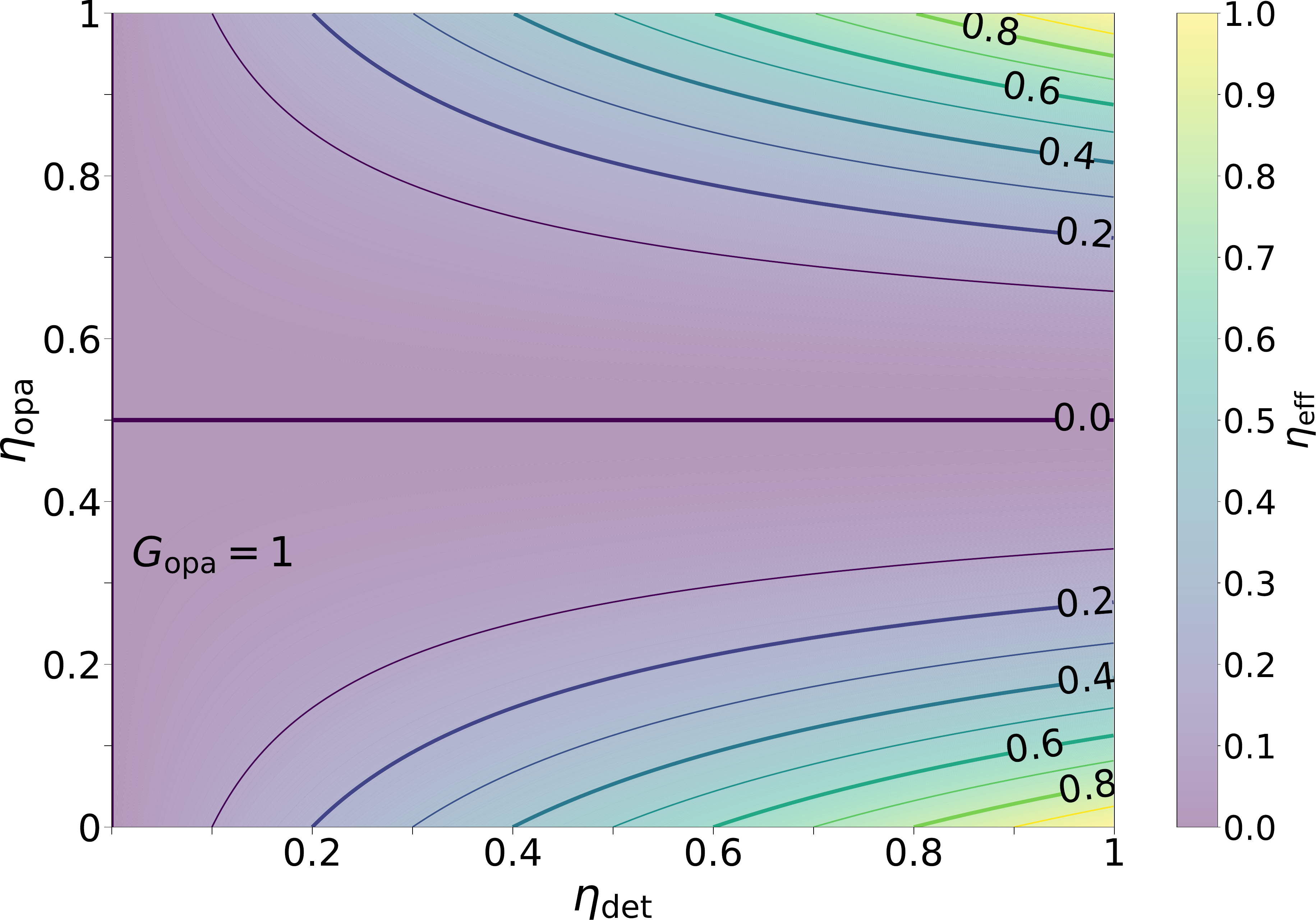}
    
  }
  \hfill
  \subfloat[\label{neff for g2=10}]{
    \includegraphics[width=0.46\textwidth]{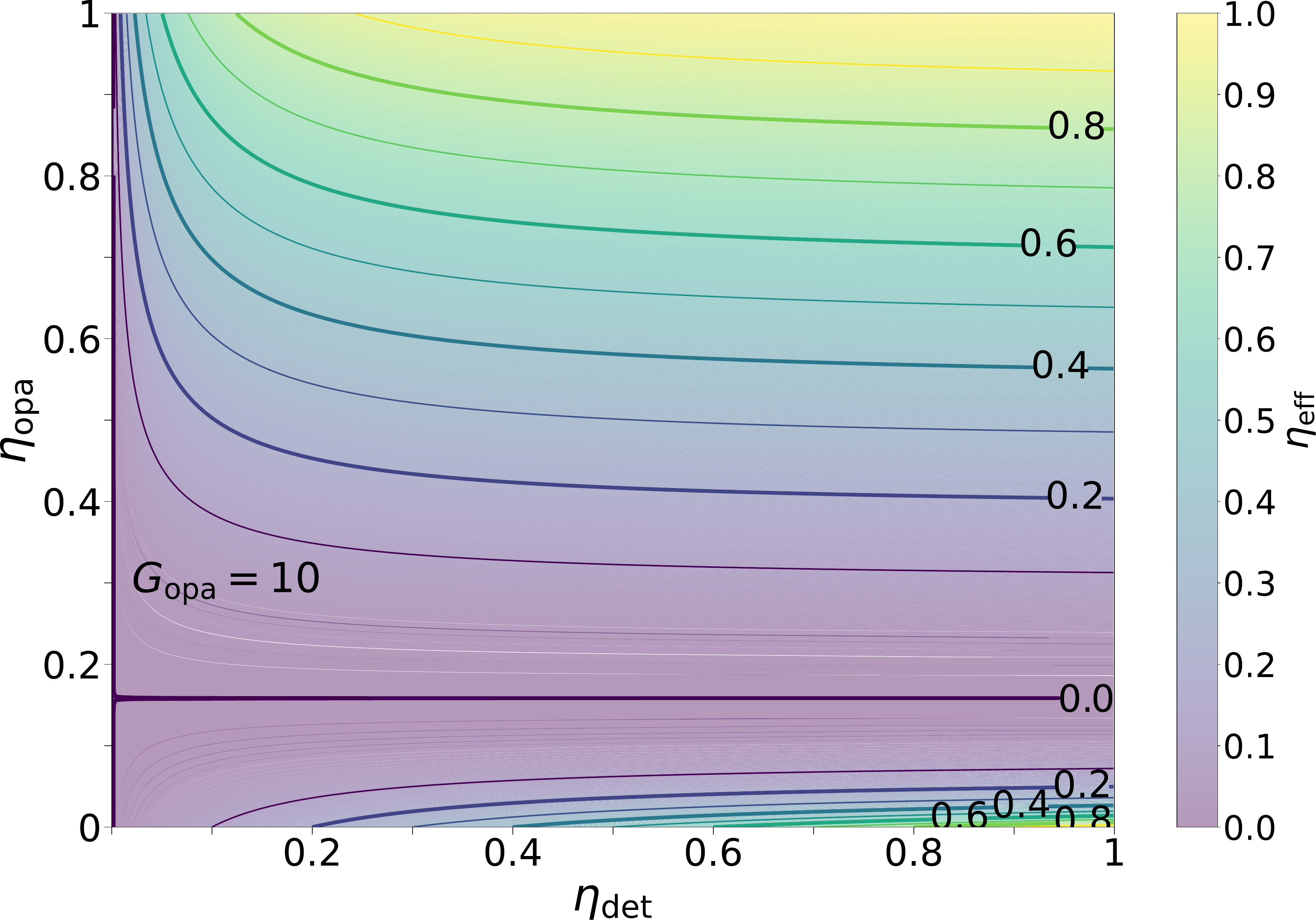}
    
  }
  \hfill
  \subfloat[\label{neff for g2= infinity}]{
    \includegraphics[width=0.46\textwidth]{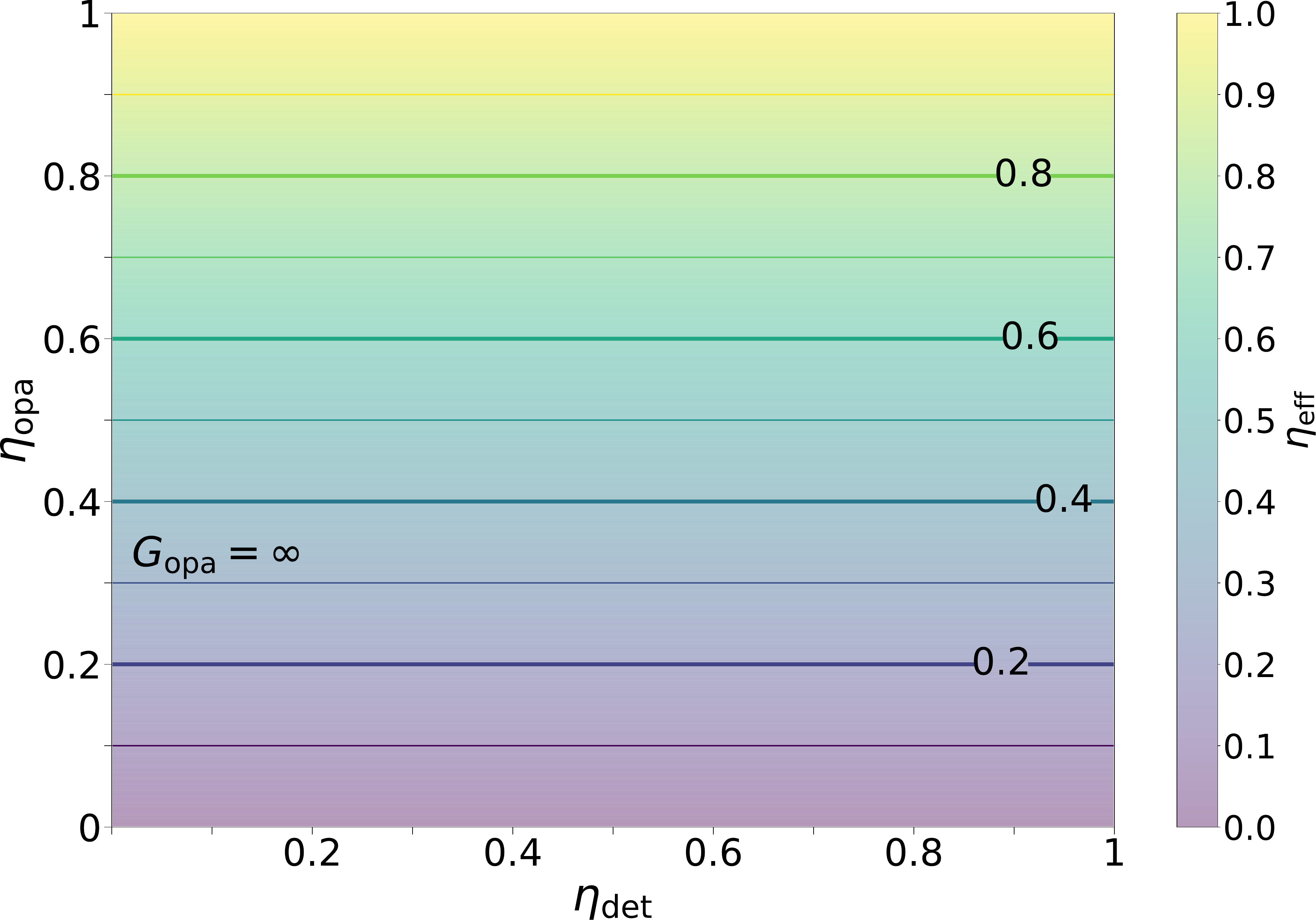}
    
  }

  \caption{Contour plots of the effective detection efficiency $\eta_{\rm eff}$ as a function of the OPA's escape efficiency $\eta_{\rm opa}$ and the detection efficiency $\eta_{\rm det}$ for three different gains $G_{\rm opa}$. To recover a high level of effective measurable squeezing, it is important to maximize the effective detection efficiency, ideally approaching unity (yellow region).} 

  \label{all neff plot}
\end{figure}

The contour plots in \Cref{all neff plot} visualize how the effective detection efficiency $\eta_{\rm eff}$ changes with the escape efficiency $\eta_{\rm opa}$ and the detection efficiency ${ \eta_{\rm det}}$. We show $\eta_{\rm eff}$ for three different plots for the gains (a) ${G_{\rm opa} = 1}$, (b) ${G_{\rm opa} = 10}$ and (c) ${G_{\rm opa} \rightarrow \infty}$. 

We first consider the simplest case with no amplification (${G_{\rm opa} = 1}$), shown in \cref{neff for g2=1} to study the effective detection efficiency. \Cref{effective efficiency} becomes:
\begin{equation}
\eta_{\rm eff}(G_{\rm opa}=1) = \eta_{\rm det} (2\eta_{\rm opa} - 1)^2 .
\label{effective efficiency for no gain}
\end{equation}
Then, $\eta_{\rm eff}$ is maximum for ${\eta_{\rm det} =1}$ and ${\eta_{\rm opa} \in \{ 0, 1 \}}$.  When the \ac{OPA} is impedance matched (${\eta_{\rm opa} = 0.5}$), the effective detection efficiency ${\eta_{\rm eff}}$ vanishes as the reflected field and the output field from the \ac{OPA} destructively interfere. The symmetry of the contours is a result of the ${(1-2\eta_{\rm opa})^2}$ term in \cref{effective efficiency for no gain}. For an over-coupled cavity (${\eta_{\rm opa} > 0.5}$), most of the input field couples into the cavity, and the output field is dominated by the transmission. In an under-coupled cavity (${\eta_{\rm opa} < 0.5}$), most of the input field is reflected and the output field is dominated by the reflection.

Figure \ref{neff for g2=10} shows $\eta_{\rm eff}$ for ${G_{\rm opa} = 10}$, which is a realistic gain for an \ac{OPA} at a wavelength of \SI{2}{\micro \metre} \cite{Georgia_squeezed_light_paper}. As the gain increases, the impedance-matched condition of the cavity reduces to ${\eta_{\rm opa} \approx 0.18}$. Because the transmitted field is amplified by the OPA, the impedance-matched case is achieved at a smaller $\eta_{\rm opa}$. 
In an over-coupled cavity with high escape efficiency where current squeezed light sources typically operate, we can mitigate large amounts of detection loss with a modest gain, as shown by the yellow region in the plot. 

For the limiting case of ${G_{\rm opa} \rightarrow \infty}$ in \cref{neff for g2= infinity}, the effective efficiency of the system $\eta_{\rm eff}$ simplifies to: 
\begin{equation}
\eta_{\rm eff}(G_{\rm opa} \rightarrow\infty) = \eta_{\rm opa} .
\label{effective efficiency for infinite gain}
\end{equation}
We can entirely mitigate detection losses for this infinite gain case because $\eta_{\rm det}$ vanishes. The effective efficiency scales proportionally only with the \ac{OPA} detection efficiency. Hence, achieving large values of $\eta_{\rm opa}$ is crucial in the implementation of the phase-sensitive amplification detection scheme. 

Our findings reveal that the effect of the detection efficiency enclosed in the $\eta_{\rm eff}$ term can be compensated by increasing the gain of the \ac{OPA}. As $G_{\rm opa}$ increases, the impact of the detection efficiency reduces. The escape efficiency of the \ac{OPA} changes from quadratic to linear as shown in \cref{all neff plot}. This can be interpreted as the gain of the \ac{OPA} mitigating the detection loss and partially mitigating the loss from the inside of the \ac{OPA} to the photodetector.

\section{Phase noise of the amplified state}
\label{section: phase noise}

A state affected by phase noise jitters around its origin in the phase space picture. At a homodyne detector, phase noise on time scales shorter than the measurement time will reduce the measured level of squeezing, while longer time scales will cause the squeezing level to drift. Assuming normally distributed fluctuations with a small standard deviation of $\tilde{\theta}$, the detected variance of the squeezed state is  \cite{Sheila_phasenoise, Oelker_phase}:
\begin{equation} 
V^{\rm conv}_{(-)}(\tilde{\theta}) = V_{(-)}^{\rm conv}\cos^2{\tilde{\theta}} + V_{(+)}^{\rm conv}\sin^2{\tilde{\theta}} .
\label{phase noise variance for single opo}
\end{equation}
 
This equation reveals that phase noise becomes more significant at high squeezing levels as the anti-squeezed quadrature is coupled into the measured squeezed quadrature. 

\begin{figure}[t!]
\centering
\subfloat[{$G_{\rm opo} = 5.2$, $G_{\rm opa} = 1$, $\tilde{\theta}_{\rm opa} = 0$} \label{Phase noise single OPO}]{\includegraphics[width=0.46\textwidth]{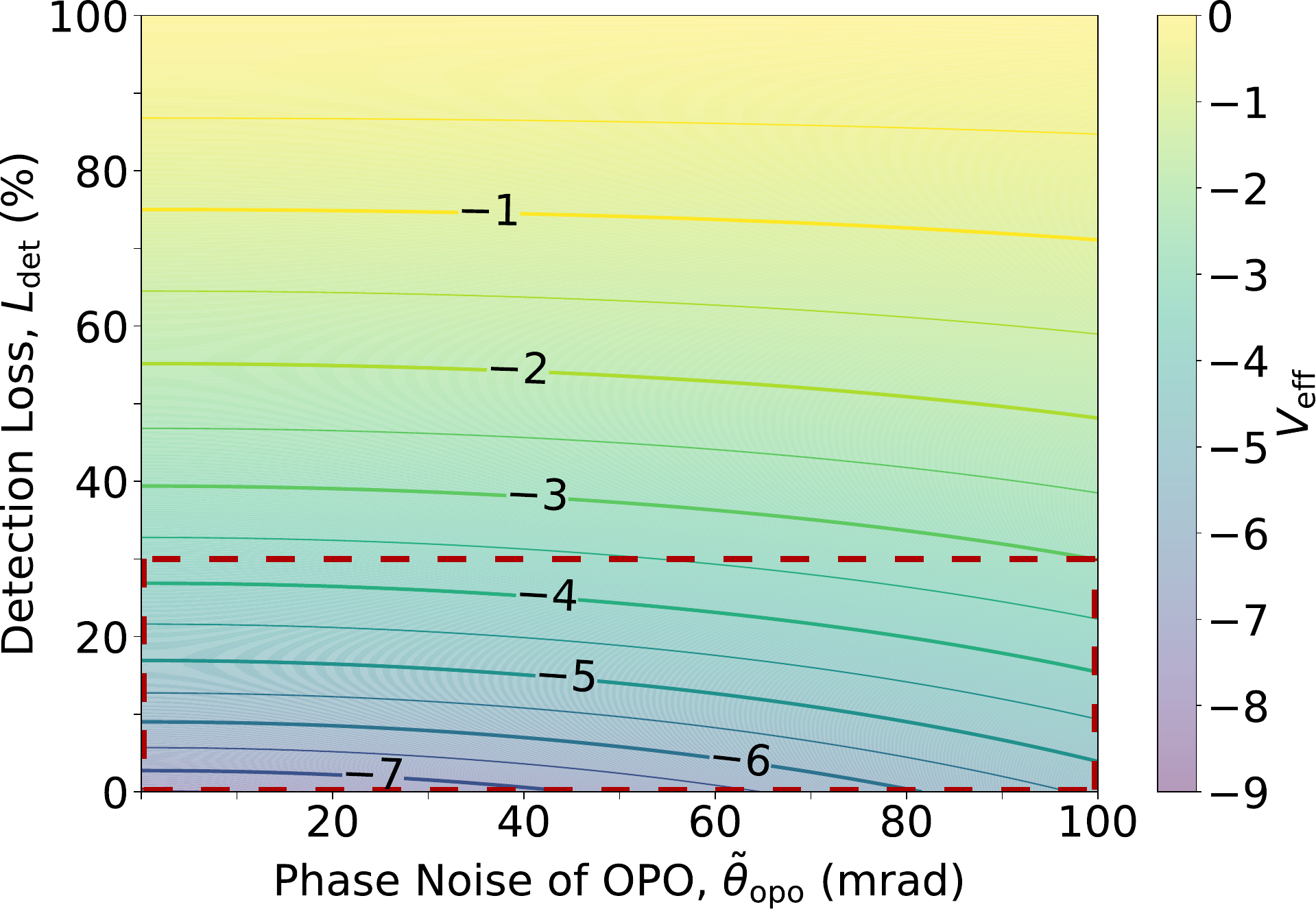}}\hfill

\subfloat[{$G_{\rm opo} = G_{\rm opa} = 5.2$, $\tilde{\theta}_{\rm opa} = 0$} \label{Phase noise in OPO 1}] {\includegraphics[width=0.46\textwidth]{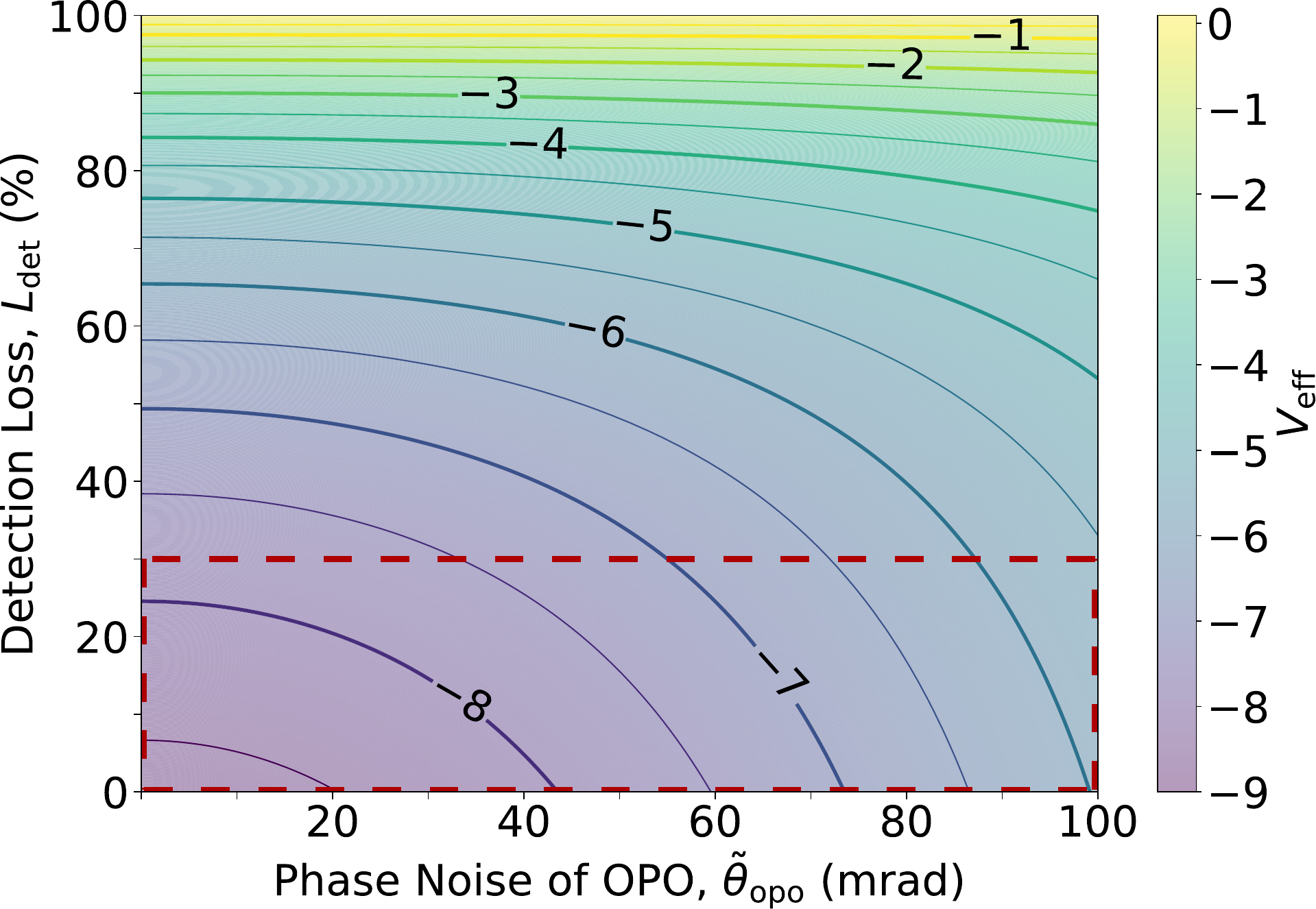}}\hfill

\subfloat[{$G_{\rm opo} = G_{\rm opa} = 5.2$, $\tilde{\theta}_{\rm opo} = 0$} \label{Phase noise in OPA 2}]{\includegraphics[width=0.46\textwidth]{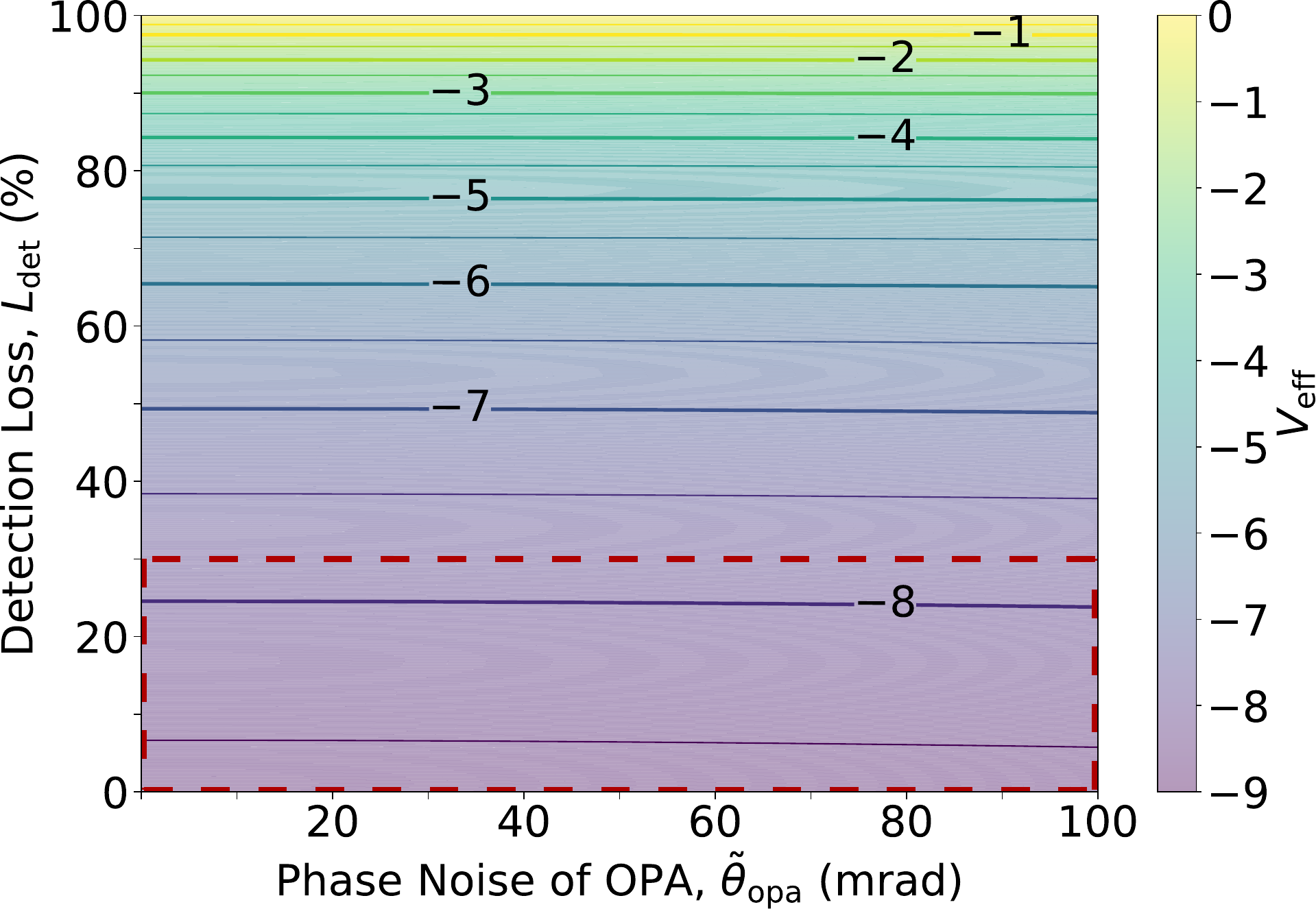}}

\caption{Contour plot of the effective measurable squeezing $V_{\rm eff}$ as a function of detection loss and phase noise, where the gain of the \ac{OPO} remains constant at $5.2$. For simplicity, only one phase noise term was considered at a time. (a) Shows the phase noise in the \ac{OPO} without amplification. (b) Shows the phase noise in \ac{OPO} with amplification. (c) Shows the phase noise in the \ac{OPA}. The red dashed area displays where the detection losses are below \SI{30}{\percent}.}

\label{Phase noise figure}

\end{figure}

We consider $\tilde{\theta}_i$, where ${i \in \{ \text{opo, opa}\}}$, for phase noise in the \ac{OPO} and \ac{OPA}, respectively. In our matrix formalism, we introduce phase noise in the form of rotational matrices \cite{Caves_twophoton,Kimble_2001}: 
\begin{equation} 
\mathbf{R}_{i} = \begin{bmatrix} \cos{(\tilde{\theta}_i}) & -\sin{(\tilde{\theta}_i)} \\ \sin{(\tilde{\theta}_i)} & \cos{(\tilde{\theta}_i}) \end{bmatrix} . 
\label{rotational matrix}
\end{equation}
The approach from the previous section is extended to include phase noise in the cascaded setup. 

The transfer matrices of the \ac{OPO} from \cref{opo input matrix,opo loss matrix} and \ac{OPA} from \cref{opa input matrix,opa loss matrix} with phase noise become:
\begin{equation} 
\mathbf{M}^{{\rm }i}_j(\tilde{\theta}_i) = \mathbf{R}_i\;\mathbf{M}^{{\rm }i}_j\;\mathbf{R}_i^{-1} , 
\label{phase noise matrix}
\end {equation}
where ${i \in \{ \text{opo, opa}\}}$ and ${j \in \{ \text{in, l}\}}$, for the input and loss port of the cavities.

The transfer functions of the system change accordingly: 
\begin{flalign}
\mathbf{TF}_{\rm in}^{\rm amp} &= \sqrt{\eta_{\rm prop} \eta_{\rm det}} \; \mathbf{R}_{\rm opa} \; \mathbf{M}_{\rm in}^{\rm opa} \; \mathbf{R}_{\rm opo} \; \notag \hspace{+1cm} \\ 
&\quad \times \mathbf{M}_{\rm in}^{\rm opo} \; \mathbf{R}_{\rm opo}^{-1} \; \mathbf{R}_{\rm opa}^{-1}, \hspace{+1cm} \\
\mathbf{TF}_{\rm lo}^{\rm amp} & = \sqrt{\eta_{\rm prop} \eta_{\rm det}} \; \mathbf{R}_{\rm opa} \; \mathbf{M}_{\rm in}^{\rm opa} \; \mathbf{R}_{\rm opo} \; \notag \hspace{+1cm} \\ 
&\quad \times \mathbf{M}_{\rm l}^{\rm opo} \; \mathbf{R}_{\rm opo}^{-1} \; \mathbf{R}_{\rm opa}^{-1}, \hspace{+1cm} \\
\mathbf{TF}_{\rm prop}^{\rm amp} & = \sqrt{(1-\eta_{\rm prop}) \eta_{\rm det}} \; \mathbf{R}_{\rm opa} \; \mathbf{M}_{\rm in}^{\rm opa} \; \mathbf{R}_{\rm opa}^{-1} ,
 \\
\mathbf{TF}_{\rm la}^{\rm amp} & = \sqrt{\eta_{\rm det}} \; \mathbf{R}_{\rm opa} \;  \mathbf{M}_{\rm l}^{\rm opa} \; \mathbf{R}_{\rm opa}^{-1} ,
\hspace{+1cm} \\
\mathbf{TF}_{\rm det}^{\rm amp} & = \sqrt{1-\eta_{\rm det}} \; \mathbf{I} . \hspace{-1cm}
\end{flalign}
Finally, the output of the \ac{OPA} is calculated from the modified transfer matrices. 

We investigate the influence of phase noise occurring at the \ac{OPO} or the \ac{OPA} for a \SI{11}{\deci \bel} squeezed state generated by the \ac{OPO} (${G_{\rm opo} = 5.2}$). Figure \ref{Phase noise figure} shows three different contour plots where the effective measurable squeezing is shown dependent on phase noise and detection loss. We highlight detection losses of $L_{\rm det} < \SI{30}{\percent}$ as a reference level as this region starts to become inaccessible to photodiodes at longer wavelengths.

The effect of phase noise occurring only at the \ac{OPO} is presented in \cref{Phase noise single OPO}, where we set the gain of the OPA to ${G_{\rm opa} = 1}$. As expected, the plot reduces to a conventional case (\cref{section: lossy_detection_of_squeezed_states}) where phase noise affects highly squeezed states the most.

\Cref{Phase noise in OPO 1} shows the effect of the same phase noise in the \ac{OPO} as in \cref{Phase noise single OPO} for a gain of ${G_{\rm opa} = G_{\rm opo} = 5.2}$. 
There is still no phase noise added to the OPA. Even for these large values of detection loss and phase noise, higher squeezing values are reached than in \cref{Phase noise single OPO} because the amplified state is less dependent on detection losses. However, the amplification causes a higher sensitivity to $\tilde{\theta}_{\rm opo}$. We observe a steeper roll-off with the increase in $\tilde{\theta}_{\rm opo}$ compared to \cref{Phase noise single OPO}. The result is consistent with the result presented in the previous section, where the degradation in the squeezing level from the \ac{OPO} is not recoverable by the amplification of the \ac{OPA}. The phase noise in the \ac{OPO} couples the anti-squeezed quadrature into the squeezed quadrature, reducing the level of squeezing entering the \ac{OPA}. 

In \cref{Phase noise in OPA 2}, we investigate the effect of phase noise in the \ac{OPA}, without phase noise in the OPO. The plot shows near-constant contour lines, demonstrating the robustness of amplified squeezed states to phase noise in our chosen measurement quadrature. As the state affected by phase noise is now close to a vacuum state (${G_{\rm opo}= G_{\rm opa}}$), the phase noise $\tilde{\theta}_{\rm opa}$ has a negligible impact for high gains $G_{\rm opa}$.

Our phase noise analysis shows that $\tilde{\theta}_{\rm opo}$ and $\tilde{\theta}_{\rm opa}$ act differently on the squeezed state. To achieve a high squeezing level, extra care has to be taken to reduce $\tilde{\theta}_{\rm opo}$. In an \ac{OPO}, the optimal gain of the system needs to be aligned with the expected phase noise to reach the best measured squeezing. Our analysis shows that this general rule is not true for the \ac{OPA} in our measurement system. We can increase the gain of the \ac{OPA} closer to the pump threshold without suffering from a large degradation of the effective measurable squeezing level.

\section{Signal-to-Noise Enhancement}
\label{section: SNR}

In this section, we extend our noise study to show how the signal is enhanced by this amplified detection scheme. We show the \ac{SNR} improvement of our amplified detection by referencing it to the conventional detection method shown in \Cref{section: lossy_detection_of_squeezed_states}.

\begin{figure}[t!]
\centering
\includegraphics[width=0.98\columnwidth]{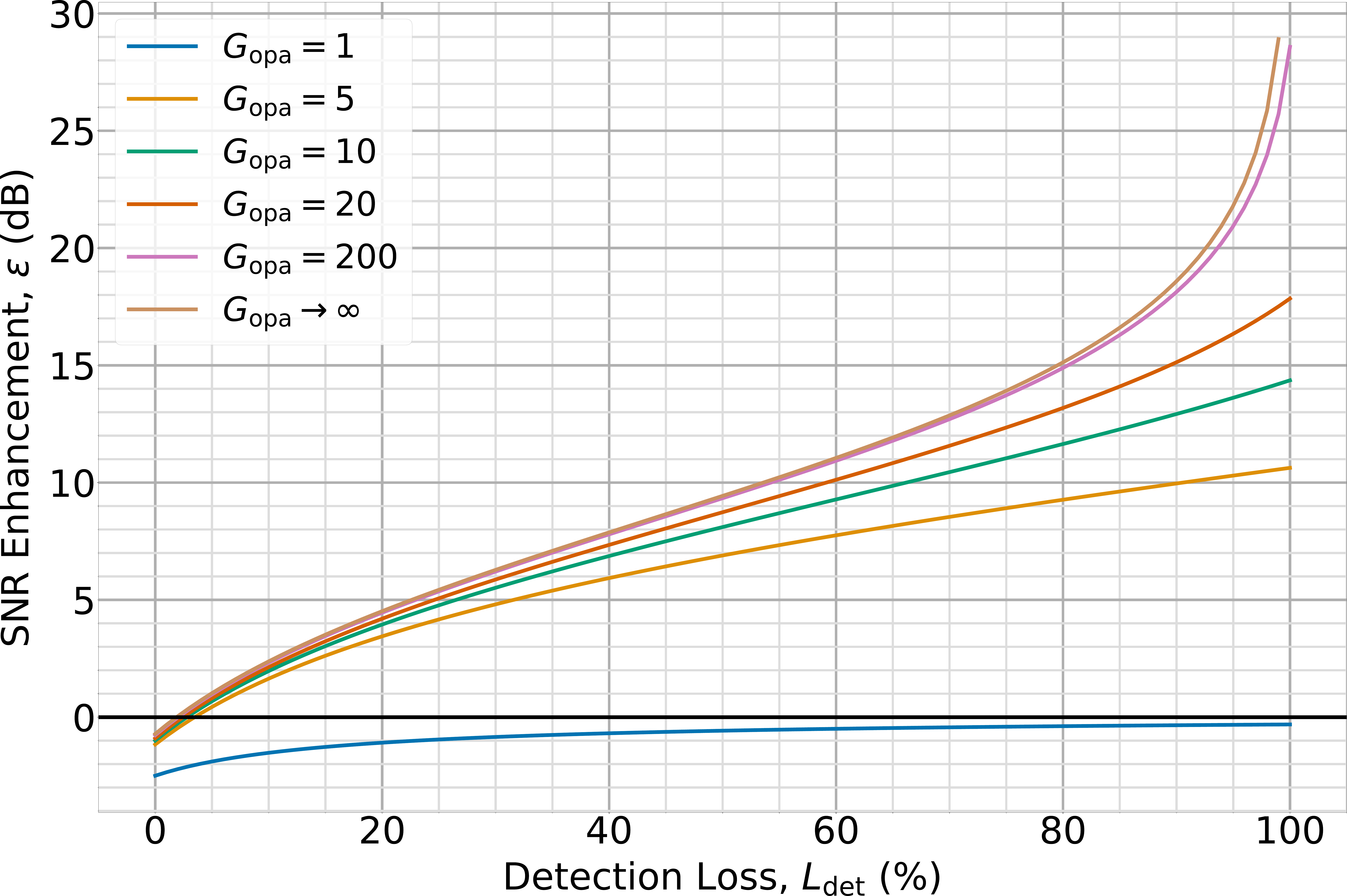}	
\caption{\label{SNR enhanced plot}The enhanced \ac{SNR} of the amplification detection scheme referenced to the conventional detection scheme plotted against the detection losses $L_{\rm det}$. }
\end{figure}

First, we calculate the \acp{SNR} of both methods individually. In the conventional detection scheme, the \ac{SNR} is given by:
\begin{equation}
{\rm SNR}_{\rm conv} = \frac{  \eta_{\rm det} \;  P_{\rm sig} } {V_{(-)}^{\rm conv}}. 
\label{SNR analysis: conventional SNR}
\end{equation}
The signal's power $P_{\rm sig}$ degrades with the detection efficiency $\eta_{\rm det}$, and the noise variance $V_{(-)}^{\rm conv}$ is described in \cref{single opo variance}. In the amplified detection method, the signal's power scales with the efficiencies $\eta_{\rm prop}$ and $\eta_{\rm det}$ but is also amplified by the \ac{OPA}. Thus, the \ac{SNR} of the amplification detection scheme is: 
\begin{equation}
{\rm SNR}_{\rm amp} = \frac{ \eta_{\rm prop} \eta_{\rm det}  \; \left( \frac{2\eta_{\rm opa}}{1-x_{\rm opa}}-1 \right)^2 P_{\rm sig} } {V_{(-)}^{ \rm amp}} .
\label{SNR analysis: amplified SNR}
\end{equation}
The \ac{SNR} enhancement $\varepsilon$ is the ratio of amplified detection and conventional detection, defined as: 
\begin{equation}
\varepsilon = \frac{ {\rm SNR}_{\rm amp}  } { {\rm SNR}_{\rm conv}  } = \frac{\eta_{\rm prop} \left( \frac{2\eta_{\rm opa}}{1-x_{\rm opa}}-1 \right)^2 V_{(-)}^{ \rm conv}}{V_{(-)}^{ \rm amp}} , 
\label{SNR analysis: enhancement}
\end{equation}
which is now independent of the signal strength. 

\Cref{SNR enhanced plot} shows the \ac{SNR} enhancement $\varepsilon$ over the detection loss $L_{\rm det}$, where the gain of the OPO is set to ${G_{\rm opo} = 5.2}$. For low detection losses (${L_{\rm det} < 5\%}$), the addition of an \ac{OPA} with non-unity escape efficiency will degrade the \ac{SNR} ($\varepsilon < 1$), regardless of the \ac{OPA} gain. For larger detection losses (${> \SI{5}{\percent}}$), the amplified detection outperforms the conventional detection ($\varepsilon > 1$). The higher the gain is of the \ac{OPA} $G_{\rm opa}$, the more significant the \ac{SNR} enhancement. Our amplified detection scheme shows the most significant \ac{SNR} enhancement when detection losses become larger. The variance quickly approaches 1 in the conventional detection, while in the amplified detection, this degradation takes longer. For a detection loss of $\SI{30}{\percent}$, it is possible to boost the \ac{SNR} by $\SI{5}{\deci \bel}$ when changing to the amplified detection method with a moderate gain of $G_{\rm opa}$.

\section{Summary and conclusions}

In this work we have used known OPO and OPA parameters, such as gain and escape efficiency, to estimate the squeezing levels under variations of phase noise for arbitrary detection losses. We showed the analytic solution for signal levels recovered with parametric amplification from any given amount of measurement loss. The solution also confirms that any squeezing lost before the amplification cannot be recovered. We then showed that phase noise in the OPA has minimal effect on the measured squeezing level as the signal is measured in the anti-squeezed quadrature of the OPA. This property enables the use of a high-gain OPA to further enhance the recovery of signals above detection losses.

A significant motivation for our model is the relatively large measurement loss that could be incurred by the next generation of gravitational wave detectors if they transition to squeezed light at a wavelength around \SI{2}{\micro\metre}. A key feature of this model is compatibility with current and proposed gravitational wave detector designs and infrastructure. Amplifier placement with regard to design considerations such as optical filter placement for frequency-dependent squeezing or for filtering higher-order spatial modes and control signals can be accommodated. 
This model is relevant for other applications of squeezed light limited by detection loss such as long-distance quantum communication protocols and output coupling from waveguides.

\begin{acknowledgments}
This research was supported by the Australian Research Council under the ARC Centre of Excellence for Gravitational Wave Discovery, Grant No. CE170100004. The authors declare no competing interests. This work has been assigned LIGO document number P2300393.

V. B. Adya would like to acknowledge the support and funding from the Swedish Research Council (VR starting grant 2023-0519 and Optical Quantum Sensing environment grant 2016-06122) and the Wallenberg Center for Quantum Technology (WACQT) in Sweden.
\end{acknowledgments}

\end{document}